\documentclass[twocolumn,showpacs,preprintnumbers,amsmath,amssymb]{revtex4}
\usepackage{dcolumn}
\usepackage{slashbox}
\usepackage{graphicx}

\newcommand{\scra}{{\mathcal A}}
\newcommand{\scrh}{{\mathcal H}}

\begin{document}

\newcommand{\be}{\begin{equation}}
\newcommand{\ee}{\end{equation}}
\newcommand{\<}{\langle}
\renewcommand{\>}{\rangle}
\newcommand{\widebar}{\overline}
\def\reff#1{(\protect\ref{#1})}
\def\spose#1{\hbox to 0pt{#1\hss}}
\def\ltapprox{\mathrel{\spose{\lower 3pt\hbox{$\mathchar"218$}}
 \raise 2.0pt\hbox{$\mathchar"13C$}}}
\def\gtapprox{\mathrel{\spose{\lower 3pt\hbox{$\mathchar"218$}}
 \raise 2.0pt\hbox{$\mathchar"13E$}}}
\def\textprime{${}^\prime$}
\def\proof{\par\medskip\noindent{\sc Proof.\ }}
\def\qed{ $\square$ \bigskip}
\def\proofof#1{\bigskip\noindent{\sc Proof of #1.\ }}
\def\half{ {1 \over 2} }
\def\third{ {1 \over 3} }
\def\twothird{ {2 \over 3} }
\def\smfrac#1#2{\textstyle{#1\over #2}}
\def\smhalf{ \smfrac{1}{2} }
\newcommand{\real}{\mathop{\rm Re}\nolimits}
\renewcommand{\Re}{\mathop{\rm Re}\nolimits}
\newcommand{\imag}{\mathop{\rm Im}\nolimits}
\renewcommand{\Im}{\mathop{\rm Im}\nolimits}
\newcommand{\sgn}{\mathop{\rm sgn}\nolimits}
\newcommand{\tr}{\mathop{\rm tr}\nolimits}
\newcommand{\diag}{\mathop{\rm diag}\nolimits}
\newcommand{\Gal}{\mathop{\rm Gal}\nolimits}
\newcommand{\mycup}{\mathop{\cup}}
\newcommand{\Arg}{\mathop{\rm Arg}\nolimits}
\def\hboxscript#1{ {\hbox{\scriptsize\em #1}} }
\def\hboxrm#1{ {\hbox{\scriptsize\rm #1}} }
\def\zhat{ {\widehat{Z}} }
\def\phat{ {\widehat{P}} }
\def\qtilde{ {\widetilde{q}} }
\renewcommand{\emptyset}{\varnothing}

\def\scra{\mathcal{A}}
\def\scrb{\mathcal{B}}
\def\scrc{\mathcal{C}}
\def\scrd{\mathcal{D}}
\def\scrf{\mathcal{F}}
\def\scrg{\mathcal{G}}
\def\scrh{\mathcal{H}}
\def\scrl{\mathcal{L}}
\def\scro{\mathcal{O}}
\def\scrp{\mathcal{P}}
\def\scrq{\mathcal{Q}}
\def\scrr{\mathcal{R}}
\def\scrs{\mathcal{S}}
\def\scrt{\mathcal{T}}
\def\scrv{\mathcal{V}}
\def\scrz{\mathcal{Z}}

\def\Z{{\mathbb Z}}
\def\R{{\mathbb R}}
\def\C{{\mathbb C}}
\def\Q{{\mathbb Q}}
\def\N{{\mathbb N}}
\def\S{{\mathbb S}}

\def\sfE{{\mathsf E}}
\def\sfT{{\mathsf T}}
\def\sfH{{\mathsf H}}
\def\sfV{{\mathsf V}}
\def\sfD{{\mathsf D}}
\def\sfI{{\mathsf I}}
\def\sfM{{\mathsf M}}
\def\sfJ{{\mathsf J}}
\def\sfP{{\mathsf P}}
\def\sfQ{{\mathsf Q}}
\def\sfR{{\mathsf R}}

\def\bone{{\mathbf 1}}
\def\bp{{\mathbf p}}
\def\bk{{\mathbf k}}
\def\br{{\mathbf r}}
\def\bm{{\mathbf m}}
\def\bn{{\mathbf n}}
\def\bl{{\mathbf l}}
\def\bq{{\mathbf q}}

\def\dbp{\tilde{d {\mathbf p}}}
\def\dbk{\tilde{d {\mathbf k}}}
\def\dbr{\tilde{d {\mathbf r}}}
\def\dbm{\tilde{d {\mathbf m}}}
\def\dbn{\tilde{d {\mathbf n}}}
\def\dbl{\tilde{d {\mathbf l}}}
\def\dbq{\tilde{d {\mathbf q}}}
\def\ddelta{\tilde{d \delta}}
\def\domega{\tilde{d \omega}}

\def\tGamma{\tilde{\Gamma}}

\def\ua{\uparrow}
\def\da{\downarrow}

\def\bv{{\bf v}}
\def\basise{{\bf e}}   
\def\basisf{{\bf f}}   
\def\startv{{\boldsymbol{\alpha}}}   
\def\endv{{\boldsymbol{\omega}}}     

\newtheorem{theorem}{Theorem}[section]
\newtheorem{definition}[theorem]{Definition}
\newtheorem{proposition}[theorem]{Proposition}
\newtheorem{lemma}[theorem]{Lemma}
\newtheorem{corollary}[theorem]{Corollary}
\newtheorem{conjecture}[theorem]{Conjecture}
\newtheorem{question}[theorem]{Question}


\newenvironment{sarray}{
          \textfont0=\scriptfont0
          \scriptfont0=\scriptscriptfont0
          \textfont1=\scriptfont1
          \scriptfont1=\scriptscriptfont1
          \textfont2=\scriptfont2
          \scriptfont2=\scriptscriptfont2
          \textfont3=\scriptfont3
          \scriptfont3=\scriptscriptfont3
        \renewcommand{\arraystretch}{0.7}
        \begin{array}{l}}{\end{array}}

\newenvironment{scarray}{
          \textfont0=\scriptfont0
          \scriptfont0=\scriptscriptfont0
          \textfont1=\scriptfont1
          \scriptfont1=\scriptscriptfont1
          \textfont2=\scriptfont2
          \scriptfont2=\scriptscriptfont2
          \textfont3=\scriptfont3
          \scriptfont3=\scriptscriptfont3
        \renewcommand{\arraystretch}{0.7}
        \begin{array}{c}}{\end{array}}

\title{Worm-type Monte Carlo simulation of the Ashkin-Teller model on the triangular lattice}
\author{Jian-Ping Lv$^{1,2,3}$, Youjin Deng$^{4}$\footnote[1]{Corresponding author: yjdeng@ustc.edu.cn},
and Qing-Hu Chen$^{3,2}$}

\address{
$^{1}$ Department of Physics, China University of Mining and Technology, Xuzhou 221116, China \\
$^{2}$ Department of Physics, Zhejiang University, Hangzhou 310027,
P. R. China\\
$^{3}$ Center for Statistical and Theoretical Condensed Matter
Physics, Zhejiang Normal University, Jinhua 321004, P. R. China \\
$^{4}$ Hefei National Laboratory for Physical Sciences at
Microscale and Department of Modern Physics, University of Science and
Technology of China, Hefei, 230027, P. R. China}
\date{\today}

\begin{abstract}
We investigate the symmetric Ashkin-Teller (AT) model on the triangular lattice in the antiferromagnetic
two-spin coupling region ($J<0$). In the $J \rightarrow  -\infty$ limit, we map the AT model onto a fully-packed
loop-dimer model on the honeycomb lattice. On the basis of this exact transformation and the
low-temperature expansion, we formulate a variant of worm-type algorithms for the AT model,
which significantly suppress the critical slowing-down. We analyze the Monte Carlo data by finite-size scaling,
and locate a line of critical points of the Ising universality class in the region $J<0$ and $K>0$,
with K the four-spin interaction.
Further, we find that, in the  $J \rightarrow -\infty$ limit,
the critical line terminates at the decoupled point $K=0$.
From the numerical results and the exact mapping, we conjecture that this `tricritical' point
($J \rightarrow -\infty, K=0$) is Berezinsky-Kosterlitz-Thouless-like and
the logarithmic correction is absent. The dynamic critical exponent of
the worm algorithm is estimated as $z=0.28(1)$ near $(J \rightarrow -\infty, K=0)$.
\end{abstract}

\maketitle

\section{Introduction}\label{ITR}

The Ashkin-Teller (AT) model is a generalization of
the Ising model to a four-component system of which each lattice site
is occupied by one of the four states~\cite{AT,Fan,Yang,AT2,AT3,Salas,per}.
In 1972, Fan ~\cite{Fan} associated each
lattice site with two Ising variables ($\sigma$,  $\tau$) and
represented the four states by the combined states $(1,1)$, $(1, -1)$,
$(-1,1)$ and $(-1,-1)$. On this basis, the reduced Hamiltonian ($k_B T\equiv 1$) of the AT
model reads
\begin{equation}
\label{Ham1}
\scrh = -\sum_{\langle i,j\rangle } \left( J_{\sigma} \sigma_i
\sigma_j+J_{\tau}  \tau_i \tau_j +K \sigma_i \tau_i \sigma_j \tau_j \right),
\end{equation}
where the sum $\langle i j \rangle $ runs over all the
nearest-neighbor pairs of spins,  $J_{\sigma}$ ($J_{\tau}$)
represents the two-spin  interaction for  $\sigma$ ($\tau$),
and $K$ is the four-spin interaction. Examples of
physical realizations of the AT model include: 1),
systems with layers of atoms and molecules adsorbed on clean surfaces--e.g., selenium
adsorbed on the Ni(100) surface ~\cite{RL1} and oxygen-on-graphite system ~\cite{RL2},
and 2), systems with layers of oxygen atoms in the CuO plane, like high-T$_{c}$
cuprate YBCO ~\cite{RL3}.

\begin{figure}
\includegraphics[width=7cm,height=5cm]{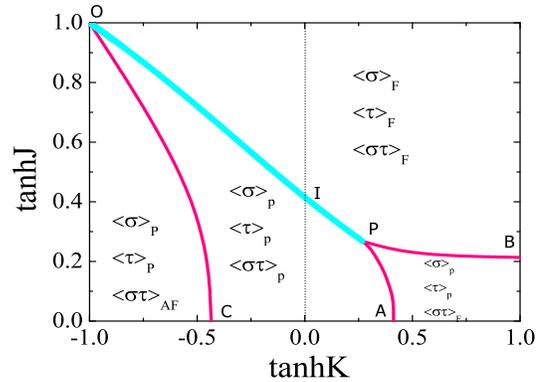}
\caption{\label{PDSQ} (Color online) Phase diagram of the AT model on the
square lattice. The `P-I-O' curve (thick cyan line) is self-dual and has continuously
varying critical exponents, separating the paramagnetic and the ferromagnetic state
in Ising variables $\sigma$, $\tau$, and $\sigma \tau$. The `P-A',  `P-B' and `O-C' lines
are commonly believed to be Ising-like, which are represented by thin magenta lines.}
\end{figure}

The AT model exhibits very rich critical behavior and plays an important
role in the field of critical phenomena. Figure~\ref{PDSQ} displays the  phase diagram
of the AT model on the square lattice for $J\equiv J_\sigma=J_\tau >0$ (we shall only consider
this symmetric case in this work).
The model reduces to two decoupled Ising systems for $K=0$, and is
equivalent to the 4-state Potts model along the diagonal line $J=K$.
The whole `P-I-O' line is critical, with continuously varying critical exponents,
and with the decoupled Ising point I and the 4-state Potts point P as two special points.
The two branches `P-A' and `P-B' are also critical, and are numerically shown to be
in the Ising universality class.
On other two-dimensional planar lattices like the honeycomb, triangular, and ka\'gome
lattices, the phase diagram of the AT model with $J \geq 0$ is similar
as Fig.~\ref{PDSQ}, except the fact that the antiferromagnetic transition line for $K<0$
may be absent on non-bipartite lattices like the triangular and the ka\'gome lattice.

In this work, we shall consider the AT model on the triangular lattice.
From the duality relation and the star-triangle
transformation, it was already found~\cite{Temperley} in 1979 that the critical
P-I-O line is described by
\begin{equation}\label{curve}
e^{-4K }=\frac{1}{2}(e^{4J}-1), \;
\end{equation}
with  $K\leq \frac{1}{4} \log 2$. Further, it can be shown that
the model on the infinite-coupling point O ($J =-K \rightarrow \infty$)
can be mapped to the critical O$(n)$ loop model with $n=2$ on the honeycomb lattice,
and the well-known Baxter-Wu model on the triangular lattice at criticality~\cite{DSS}.
In the limit $J =K \rightarrow -\infty$, the model is equivalent to
the 4-state Potts antiferromagnet at zero temperature, which is also critical.
In the limit $J=0, K \rightarrow -\infty$, the AT model
reduces to the zero-temperature Ising antiferromagnet in
variable $\sigma \tau$; the same applies to
the limit $K =0, J \rightarrow -\infty$  for the two decoupled Ising
variables $\sigma$ and $\tau$.
Phase transition of the triangular-lattice Ising antiferromagnet is absent at finite temperature, and
at zero temperature the system has non-zero entropy per site~\cite{wannier,frustate}.
The pair correlation on any of the three sublattices of the triangular lattice
decays algebraically as a function of distance, and
the associated magnetic scaling dimension is $X_h=1/4$ ~\cite{stephenson}.

On the square lattice, the phase diagram
for $J<0$ is the symmetric image of Fig.~\ref{PDSQ} with respect to the $K$ axis ($J \rightarrow -J$),
arising from the bipartite property.
However, to our knowledge, the phase diagram of the
AT model is still unknown on the triangular and the ka\'gome lattice with $J <0$.
Clearly, the Ising critical line $P-A$ should continue into the region $J<0, K>0$,
albeit it remains to be explored how this extension looks like.
Due to the absence of exact result, we will apply Monte Carlo method
and the finite-size scaling theory.
Monte Carlo simulation of the triangular AT model is challenging for large negative coupling $J<0$,
arising from the so-called geometric frustration.
Antiferromagnetic coupling $J<0$ means that the neighboring Ising spins prefer to be anti-parallel.
However, such a preference cannot be satisfied for all of the three neighboring pairs on
any elementary triangular face. One can at most have two antiferromagnetic pairs.
For such a frustrated system, most Monte Carlo simulation suffers
significantly from critical slowing-down.
In fact, as $J \rightarrow -\infty$, the Metropolis and the Swendsen-Wang-type cluster
algorithm are found to be non-ergodic~\cite{Salas,SwendsenWang,ClusterTri,frustration1}.
Recently, worm-type algorithms with the so-called rejection-free was developed for the antiferromagnetic Ising model on
the triangular lattice and other systems~\cite{improved,QQLiu}.
This algorithm has been proved to be ergodic at zero
temperature and only suffers from minor critical slowing-down.
The rejection-free worm algorithm can be extended to the AT model,
albeit the efficiency is limited for nonzero $K$ in the zero-temperature limit $J \rightarrow - \infty$.

The outline of this paper is as follows. Section~\ref{LT}
describes the partition sum of the AT model as well as an exact mapping
to a fully-packed loop-dimer (FPLD) model in  $J \rightarrow - \infty$ limit.
A variant of worm-type algorithms is developed in Sec.~\ref{WORM}.
The numerical results are presented in Sec.~\ref{results}.
In Sec.~\ref{worm dynamic data} we investigate the dynamic critical behavior of
one of the worm algorithms.
A discussion is given in Sec.~\ref{discussion}, including the phase diagram on the ka\'gome lattice.

\section{Model and exact mapping}\label{LT}
\subsection{Low-temperature expansion of the AT model}\label{LT1}
\begin{figure}
\includegraphics[width=8cm]{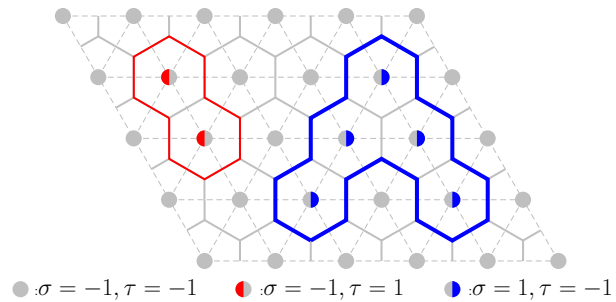}
\caption{\label{LTEG}
   (Color online) A spin configuration of the AT model on the triangular lattice and the corresponding
   LT-expansion graph on the honeycomb lattice. Thick blue line represents blue bond;
   thin red line denotes red bond; the same below.}
\end{figure}
Instead of directly updating the spins, the worm-type algorithms~\cite{worm1,worm2} for the Ising model
simulate the graphical representation which can be the high- and the low-temperature
expansion graphs. The worm methods in Refs.~\cite{worm1,worm2} can be generalized to
the graphical expansion of the AT model. In the following, we shall use the low-temperature (LT)
expansion, defined on the dual lattice of the triangular lattice---the honeycomb lattice.
Given a spin configuration $\{\sigma, \tau \}$, for each pair of nearest-neighboring
vertices $(i,j)$, one places on its dual edge:
\begin{itemize}
\item nothing if $\sigma_i=\sigma_j, \tau_i =\tau_j$,
\item a red occupied bond if $\sigma_i = \sigma_j, \tau_i \neq \tau_j$,
\item a blue occupied bond if  $\sigma_i \neq \sigma_j, \tau_i =\tau_j$,
\item a red and a blue bond if $\sigma_i \neq \sigma_j, \tau_i \neq \tau_j$.
\end{itemize}
In other words, depending on the associated pair of spins on the triangular lattice, an edge on the honeycomb
lattice can be in one of the four states: vacant, red, blue, and red+blue. An example is shown in
Fig.~\ref{LTEG}. Since the coordination number is 3 for the honeycomb lattice, the red and blue
bonds form a series of disjointed loops in red and blue color, respectively.
Note that the red and the blue loops are allowed to share common edges.
In this way, a spin configuration on the triangular lattice
is mapped onto a loop configuration on the honeycomb lattice, while a loop configuration
corresponds to 4 spin configurations~\footnote{this is not precisely correct
for torus geometry, where a loop configuration can correspond to no spin configuration.},
which are related to each other by globally flipping the $\sigma$ or/and $\tau$ Ising spins.
Let $|E_r|$, $|E_b|$, and $|E_{r+b}|$ be the number of red, blue, and red$+$blue bonds,
the partition sum of the AT model can be written as (up to an unimportant factor)
\begin{equation}
  \scrz_{\rm AT} = \sum_{\{ \scrl \}} X_r^{|E_r|} X_b^{|E_b|} X_{r+b}^{|E_{r+b}|} \; ,
\label{partition_lt1}
\end{equation}
where the summation  $\{ \scrl \}$ is over all loop configurations.
From the mapping, one can obtain the relative statistical weights as
\begin{equation}
  X_r=X_b=e^{-2J-2K} \mbox{   and   } X_{r+b}=e^{-4J} \; .
\label{prob1}
\end{equation}

\begin{figure}
\includegraphics[width=9cm]{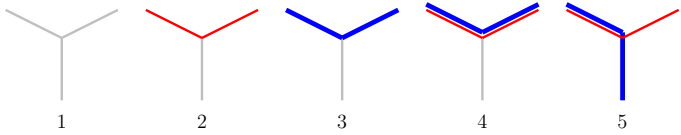}
\caption{\label{sts} (Color online)  Vertex states in the LT-expansion graph of the AT model.}
\end{figure}

One can further describe the AT model in the language of the vertex states,
which will serve as the basis for the formulation of the worm-type algorithms in this work.
In the loop configurations, all the vertices must have an even number of incident
red (blue) bonds.  Accordingly, only the 5 types of vertex states in Fig.~\ref{sts}
exist, where the states are unchanged under spatial rotations.
Simple calculations yield the statistical weights as
\begin{eqnarray}
  W_1 &=& 1\; ,       \hspace{13mm} W_2 = W_3 = e^{-2J-2K}\; , \nonumber \\
  W_4 &=& e^{-4J}\; , \mbox{ and }  W_5 =       e^{-4J-2K} \; .
  \label{vertex_weight_AT}
\end{eqnarray}
Let $|V_i|$ be the number of vertices at state $i$ with $i=1,2,3,4,5$, the partition sum of the
AT model can be written as (up to a constant)
\begin{equation}
  \scrz_{\rm AT} = \sum_{\{ \scrv \} } \prod_{i=1}^5 W_i^{|V_i|} \; ,
\label{partition_lt2}
\end{equation}
where the summation $\{ \scrv \}$ is over configurations with vertex states in Fig.~\ref{sts}.

\subsection{Exact mapping in the $J \rightarrow - \infty$ limit}\label{LT2}

\begin{figure}
\includegraphics[width=8cm]{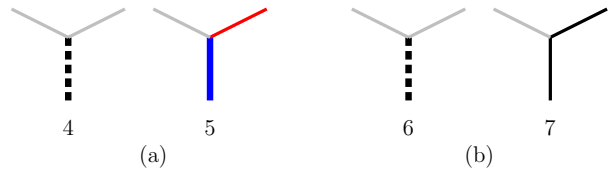} \hspace{6mm}
\caption{\label{fp_ld}
(Color online) (a), State-4 and -5 after the redefinition of the edge states. (b),
Vertex states in the FPLD model. The dashed black line represents dimer.}
\end{figure}

Given a finite four-spin coupling $K$, when the antiferromagnetic coupling $J$
becomes stronger and stronger, more and more vertices will be at state-4 and -5 in Fig.~\ref{sts},
because $W_4 \sim W_5 \propto \exp (-4J)$ increases faster than $W_1, W_2, W_3$, as seen
from Eq.~(\ref{vertex_weight_AT}).
In the $J \rightarrow - \infty$ limit, only state-4 and -5 survive. We can then redefine
the edge states in state-4 and -5 as following.
The empty edge is replaced by a `dimer', while the `blue+red' edge is
regarded as empty; namely, the edge is now at state:  empty, dimer, red, or blue.
As a result, state-4 and -5 become those in Fig.~\ref{fp_ld}(a).
One observes that the occupied bonds at state-5 form a series of disjointed loops;
these loops are now constructed by bonds {\it alternatively} in color red and blue.
Further, one notes that the color-degree freedom can be simply integrated out, and each loop gains
a statistical-weight factor 2. Without the color information, the edge is at state: empty, dimer,
or bond, and the vertex states reduce to those in Fig.~\ref{fp_ld}(b), where new labels `6' and `7'
are used.  The statistical weights are
\begin{equation}
  W_6 = 1 \; , \hspace{6mm} W_7 = e^{-2K} \; .
\end{equation}

On this basis, the partition sum of the AT model in the $J \rightarrow - \infty$ limit can be written as
\begin{equation}
  \scrz_{\rm FPLD} = \sum_{\{\scrv \}} n^{\ell} W_7^{|V_7|}  \; , \hspace{6mm} (n=2)
\label{loop_dimer}
\end{equation}
where the summation is over configurations with all vertex states in Fig.~\ref{fp_ld}, and
$\ell$ is the number of loops.
We shall refer to the model defined by Eq.~(\ref{loop_dimer}) and Fig.~\ref{fp_ld}
as the $n$-color FPLD model.

Note that, for finite $K$, the loops in the FPLD model are `dilute' due to
the presence of state-6.
However, in the $K \rightarrow - \infty$ limit, only state-7 survives, and
one obtains the mapping between the triangular 4-state antiferromagnet
at zero temperature  and the honeycomb $n=2$ fully-packed loop model.
For $K \rightarrow \infty$,
the model reduces to the fully-packed dimer model, which is equivalent to the triangular
Ising antiferromagnet at zero temperature.

We conclude this subsection by mentioning that the FPLD model is
very similar to the honeycomb O$(n)$ loop model~\cite{Nienhuis}. The difference is that
in the former the vertices off the loops are paired up by dimers,
while not in the latter. Namely, the configuration space for the FPLD model
is a subspace in the O$(n)$ loop model. Albeit it remains to be explored whether or not
the two models are in the same universality class, it is not surprising if this turns out to be the case.

\section {Worm Algorithms}\label{WORM}

The worm algorithm for the high-temperature expansion graphs of the Ising model was
first formulated by Prokof'ev and Svistunov~\cite{worm1}, and the dynamic critical
behavior was studied in Ref.~\cite{worm2}.  Recently, Wolff provided a worm-type simulation
strategy for O(N) sigma/loop models~\cite{wolff}.
The underlying physical picture of the worm method is beautifully simple:
enlarge the state space of the to-be-simulated model, define an
extended model, and simulate the system by a local algorithm.

\subsection {Worm algorithm for finite $J$}\label{standard}

Let us now generalize the worm method in Refs.~\cite{worm1,worm2}
to the AT model in the language of the vertex states,
defined by Eq.~(\ref{partition_lt2}) and Fig.~\ref{sts}.

\begin{figure}
\includegraphics[width=8cm]{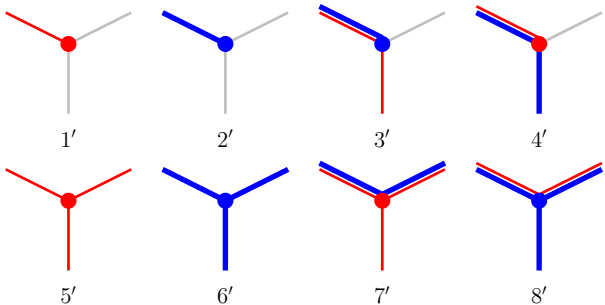}
\caption{\label{ustsA} (Color online) Additional vertex states in the worm method for the AT model.
The red (blue) filled circle denotes a defect in red (blue) vertex configuration.}
\end{figure}

{\em Enlarge the state space.} We first introduce new vertex states by
deleting from (or adding to) the states in Fig.~\ref{sts} a red or blue bond.
This leads to the 8 additional vertex states in Fig.~\ref{ustsA}.
The state space is then enlarged such that a configuration has
{\em a pair} or {\em none} of vertices at states in Fig.~\ref{ustsA}.
Such a pair of vertices are named `defects' and denoted as $(u,v)$.
Accordingly, the state space can be divided into two subspaces: one without
defect ($u=v$) and the other with two defects ($u \neq v$); we shall
refer to them the M (measuring) and W (worm) sector, respectively.
A careful check yields
that the pair of defects in the W sector must be connected via a string of red or blue
occupied bonds. Namely, $u$ and $v$ are either both at states $\{1', 4', 5', 7' \}$
or $\{2', 3', 6', 8' \}$ in Fig.~\ref{ustsA}.
For the later convenience, we let $u, v$ be ordered as $u \leftarrow v$, and thus
the interchange $(u \leftrightarrow v)$ would lead to a different configuration for $u \neq v$.

{\em Define the extended model.} With the inclusion of the defects and
the vertex states in the W sector, a configuration can now be
completely specified by its vertex
states $\{ \scrv \}$, and the ordered pair of defects $(u,v)$. The partition sum of the
extended model can be separated into two parts.
The part in the M sector is defined as
\begin{equation}
  \scrz_M = \scrz_{\rm AT} = \frac{1}{V} \sum_{\{ \scrv, u, v\}} \delta_{u = v}
  \prod_{i=1}^5 W_i^{|V_i|} \;,
\label{weight_M_sector}
\end{equation}
where $V$ is the volume of the system and $\delta$ is the Kronecker delta function.
The summation $\{ \scrv \}$ is over vertex-state configurations with
states in Fig.~\ref{sts} and coordinations $(u,v)$.
Factor $1/V$ accounts for the summation of $u=v$ over the whole lattice.
Similarly, the part of the partition sum in the W sector can be defined by
\begin{equation}
  \scrz_W = \frac{1}{V} \sum_{\{ \scrv, u, v\}} \delta_{u \neq  v}
  \prod_{i=1}^{5} W_{i}^{|V_i|}\prod_{j=1}^{8}  W_{j'}^{|V_{j'}|} \; ,
\label{weight_W_sector}
\end{equation}
where the summation $\{ \scrv \}$ is over configurations with two vertex states in Fig.~\ref{ustsA}
and all other in Fig.~\ref{sts}, and $W_{j'}$ are the statistical weights for states in Fig.~\ref{ustsA}.
The extended model can then be defined as
\begin{equation}
  \scrz_{\rm worm} = \scrz_M + \xi_w \scrz_W \; ,
  \label{extended_model}
\end{equation}
with $\xi_w >0$ a constant factor controlling the relative weight in the M and the W sector.

For a complete definition of the extended model, the statistical weights $W_{j'}$ for states
in Fig.~\ref{ustsA} should have a definite value.
It is natural that they are defined in accordance with the edge states, which lead to
\begin{eqnarray}
  W_{1'} = W_{2'} &=& e^{-J-K}   \; , \hspace{3mm} W_{3'} = W_{4'} = e^{-3J-K} \; , \nonumber \\
  W_{5'} = W_{6'} &=& e^{-3J-3K} \; , W_{7'} = W_{8'} = e^{-5J-K}  \; .
  \label{vertex_weight_W}
\end{eqnarray}

{\it Formulate the worm algorithm.} One can now use any valid algorithm to
simulate the model defined by Eq.~(\ref{extended_model}). Since a configuration is specified by
the ordered triplet of parameters $(\scrv,u,v)$, an update can be acted on the vertex states
$\scrv$ and/or the locations of defects $(u,v)$.
The worm strategy is to randomly move $u$ and/or $v$ around the lattice and
update $\scrv$ by changing the edge states during the biased random walk.
Suppose that $u \neq v$ are in red (in the W sector).
As $u$ moves to a neighboring vertex $u_n$,
the edge $(uu_n)$ state will be symmetrically updated: a red bond is placed (deleted) if it is absent (present).
In this way, state at $u$ will be back in Fig.~\ref{sts} after $u$ moves $u_n$.
Accordingly, the number of defects remains unchanged if $v \neq u_n$ or becomes zero
if $v = u_n$. This accounts for a step of random walk in the W sector or from the W to the M sector.
For the case $u=v$, by the symmetric update of edge state, one will generate a pair
of defects which can be either in red or blue. Therefore, one never introduces more than two defects.
The parameter $\xi=1$ is set in this work, and a version of the worm algorithm reads ({\it Algorithm 1})
\begin{enumerate}
\item If $u=v$, randomly choose a new vertex $u'$ and set $u=v=u'$.
      Equally choose color red or blue for the to-be-proposed defects; say red.
\item Interchange $u \leftrightarrow v$ with probability $1/2$.
\item Randomly choose one neighboring vertex $u_n$ of $u$. Propose to move $u \rightarrow u_n$.
\item Propose to symmetrically update the edge-$uu_n$ state:
      red $\leftrightarrow$ vacant and blue $\leftrightarrow$ red+blue.
\item Accept the proposal with probability
      \begin{equation}
       \scrp_a = \min \left[1, (W^{(a)}_{u} W^{(a)}_{u_n})/(W^{(b)}_{u} W^{(b)}_{u_n})\right] \; , \nonumber
      \end{equation}
      according to the Metropolis-Hasting scheme. The superscript $(b)$ and $(a)$
      means ``before'' and ``after update'', respectively. The statistical weights
      are given Eqs.~(\ref{vertex_weight_AT}) and (\ref{vertex_weight_W}).
\end{enumerate}
Monte Carlo simulation of the AT model consists of repetition of these steps.
The detailed balance at each step is straightforward since the algorithm
is just a Metropolis-type update. If one regards the connected pair of
`defects' as a worm, the above steps mimic the crawling of the worm on the lattice.
This is responsible for the terminology `worm'.

{\it Measurement.} Measurement can take place either in the whole enlarged state space
or in the M subspace. For the high-temperature graph of the Ising model,
it can be shown that the partition sum of the extended model is related to
the Ising model as $\scrz_{\rm worm} = \chi \scrz_{\rm Ising}$,
where $\chi$ is the magnetic susceptibility. Thermodynamic quantities
can be measured in the enlarged configuration space. Nevertheless,
if one is only interested in the original system, it is sufficient to
sample in the M sector. This would define a Markov subchain with a coarse unit of Monte Carlo step
between two subsequent configurations in the M sector. The detailed balance is
clear since it is satisfied in each basic step in {\it Algorithm 1}.

{\it Improved version.} As mentioned earlier, state-4 and -5 (Fig.~\ref{sts})
would dominate in the M sector as $J \rightarrow - \infty$;
analogously, only state-$7'$ and -$8'$ (Fig.~\ref{ustsA}) survive in the limit, as
seen from  Eq.~(\ref{vertex_weight_W}).  This implies that,
as soon as both $u$ and $v$ are at state-$7'$ and $-8'$, they will be frozen there forever,
and thus {\it Algorithm 1} becomes non-ergodic.

The same difficulty occurs for the worm simulation of the triangular Ising antiferromagnet
at zero temperature.  A rejection-free technique was introduced~\cite{improved,QQLiu}
to overcome such a problem, based on the observation that the detailed balance in the
coarse step does not require the detailed balance in each basic step in the W sector.
Let $u_n \; (n=1,2,3)$ denote the neighbouring vertices of $u$ and
$p_n$ be the probability that $u$ moves to $u_n$ in {\it Algorithm 1},
the probability for $u$ to be unmoved is $p_0=1-(p_1+p_2+p_3)$.
The absorbing problem of $(u,v)$ at state-$7'$ and -$8'$ is reflected by
$p_0 \rightarrow 1$ as $J \rightarrow - \infty$.
In the W sector, one can explicitly set {\it zero} for the probability that $u$ remains unmoved,
and defines the new transition probabilities $p'_n$ as
\begin{eqnarray}
  \frac{p'_1}{p_1} = \frac{p'_2}{p_2} = \frac{p'_3}{p_3} \; , \nonumber \\
  p'_0=1-(p'_1+p'_2+p'_3) =0 \; .
\end{eqnarray}
The details can be found in Refs.~\cite{improved,QQLiu}.

The absorbing problem can also be solved in the present formulation of the worm algorithm.
Actually, the  absorbing problem is somewhat `artificial' here, since it arises from
the particular assignment of the statistical weights to
states in Fig.~\ref{ustsA} by Eq.~(\ref{vertex_weight_W}). There is no reason, however, why one should
use Eq.~(\ref{vertex_weight_W}) if only the original AT model~(\ref{partition_lt2}) is of interest.
The absorbing problem simply dissolves if the statistical weights are given by
\begin{eqnarray}
  W_{1'} = W_{2'} &=& e^{-2J-2K} \; , W_{3'} = W_{4'} = e^{-4J} \; , \nonumber \\
  W_{5'} = W_{6'} &=& e^{-2J-2K} \; , W_{7'} = W_{8'} = e^{-4J-2K}\label{vertex_weight_W1} \;.
\end{eqnarray}
Other definitions are possible.

\subsection {Worm algorithm for $J \rightarrow - \infty$}\label{IWA}

{\it Algorithm 1} using Eq.~(\ref{vertex_weight_W1}) is found to be efficient
in most of the region with $0>J>-\infty$ and for small $K$ in the $J \rightarrow -\infty$ limit.
In this limit, the efficiency significantly drops as $K$ deviates from $0$.

\begin{figure}
\includegraphics[width=8cm]{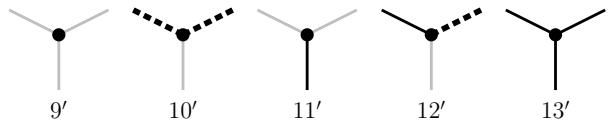}
\caption{\label{fp_ld_W}
   (Color online) Vertex states in the W sector for the FPLD model. The black filled circle denotes a defect.}
\end{figure}

Hereby we shall make use of the exact mapping of the AT model onto the
FPLD model~(\ref{loop_dimer}) and formulate another version of the worm algorithm.
Following the same procedure in the above subsection, we first introduce
5 additional states in Fig.~\ref{fp_ld_W}.
The partition sum in the M sector is defined as
\begin{equation}
  \scrz_M = \scrz_{\rm FPLD} = \frac{1}{V} \sum_{\{ \scrv, u, v\}} \delta_{u = v} n^\ell W_7^{|V_7|} \; ,
\end{equation}
with $n=2$. Again, the summation is over configurations with states in Fig.~\ref{fp_ld}
and over the location of $u=v$. The partition sum in the W sector is given by
\begin{equation}
   \scrz_W = \frac{1}{V} \sum_{\{ \scrv, u, v\}} \delta_{u \neq  v} n^{\ell}
     W_7^{|V_7|}\prod_{j=9}^{13}  W_{j'}^{|V_{j'}|} \; .
\end{equation}
The extended model is defined by Eq.~(\ref{extended_model}).

The formulation of the worm algorithm follows the standard strategy in the above subsection,
except that the edge-state update should take a different scheme. Let $e=0, 1, 2$ denote
the edge-$e$ state `empty', `bond', and `dimer', respectively, and define the module-3
summation rule as $\mod_3(e+\Delta e)$ with $\Delta e=1,2$.
As moving $u \rightarrow u_n$, one randomly chooses $\Delta e =1$ or 2 and propose to
update the edge-$uu_n$ state as $\mod_3(e+\Delta e)$. In other words, an `empty' edge
is proposed to randomly become a `bond' or a `dimer'; `dimer' is to be `empty' or `bond';
and `bond' is to be `empty' or `dimer'. However, not all the proposals will generate
a valid configuration that has {\it at most} two states in Fig.~\ref{fp_ld_W} and
the others in Fig.~\ref{fp_ld}. For instance, (1), in the M sector, when $u=v$ is at state-$7$ and
the empty edge is proposed to become a dimer, the resulting vertex state at $u$ will not be in
Fig.~\ref{fp_ld_W}; (2), in the W sector, when $u$ is at state-$9'$ and the proposal is
$e=0 \rightarrow e=1$, this would yield state-$11'$ at $u$ which is not in Fig.~\ref{fp_ld}
as required. A proposal would be rejected if it leads to an invalid configuration.
On this basis, a version of the worm algorithm can be formulated as ({\em Algorithm 2})
\begin{enumerate}
\item If $u=v$, move it to a randomly chosen vertex.
\item Same as in {\it Algorithm 1}.
\item Same as in {\it Algorithm 1}.
\item Randomly choose $\Delta e = 1$ or $2$, and propose to update the edge-$uu_n$ state as $e_{uu_n}
      \rightarrow \mod_3(e_{uu_n}+\Delta e)$. The proposal will be rejected if it yields
      \begin{itemize}
	\item for $u=v$,                        $V_v$ or $V_{u_n}$ $\not\in \{9',\cdots, 13' \}$ in Fig.~\ref{fp_ld_W};
	\item for $u\neq v$ and $v \neq u_n$,   $V_u$ $\not\in \{6,7\}$ in  Fig.~\ref{fp_ld} or $V_{u_n}$
                           	                $\not\in \{9',\cdots, 13' \}$;
        \item for $u\neq v$ and $v = u_n$,      $V_u$ or $V_{u_n} $  $\not\in \{6,7\}$.
      \end{itemize}
      In this case, step-5 will be skipped. Symbol $V_u$ represents the vertex state at $u$.
\item Accept the update with probability
      \begin{equation}
       \scrp_a= \min \left[1, n^{\Delta_\ell} \; (W^{(a)}_u W^{(a)}_{u_n})/(W^{(b)}_{u}W^{(b)}_{u_n})\right] \; ,
       \nonumber
      \end{equation}
      where $\Delta \ell$ denotes the change of the loop number in the update.
      We remind that the constant $\xi_w$ in Eq.~(\ref{extended_model}) is set $\xi_w=1$.
\end{enumerate}
Simulation consists of repetition of these steps, and the measurement is taken in the M sector.

A practically important matter for implementing {\it Algorithm 2} is that a non-local query is needed
to calculate the loop-number difference $\Delta \ell$. We shall follow the simultaneous breadth-first searching
technique and the trick to avoid as much as possible queries, as described in Ref.~\cite{QQLiu}.

More importantly, one can apply the so-called coloring method to avoid altogether the need for such global queries
for $n \geq 1$. The key ingredient of the coloring method is the trivial identity $n = 1+(n-1)$
for the statistical weight $n$ of each loop. One can introduce an auxiliary variable $c =0, 1$
and rewrite the identity as
\begin{equation}
  n = \sum_{c=0,1} [1 \delta_{c,0} + (n-1) \delta_{c,1}] \; .
\end{equation}
The variable $c$ is generally referred to as the coloring variable, and $c=0$ (1) is said `active' (`inactive') .
See Refs.~\cite{QQLiu} for details.
In practise, the coloring variable is assigned to each vertex in the M sector as ({\it Coloring assignment})
\begin{enumerate}
  \item Set all vertices off loops be active ($c=0$).
  \item Independently for {\it each} loop, choose $c=0$ with probability $1/n$
        and $c=1$ with probability $(1-1/n)$, and assign it to all the vertices on the loop.
\end{enumerate}
On the basis of the {\it Coloring assignment}, the whole lattice $G$ is divided into the active sublattice $G_a$
and the inactive sublattice $G_i$. In $G_a$ the vertices are active and the edges connect two active vertices;
in $G_i$ the vertices are inactive and the edges connect two inactive vertices. The edges connecting one
active and one inactive vertex form the boundaries separating $G_a$ and $G_i$. We state that,
conditioning on this decomposition, the vertex-state configuration on the induced sublattice $G_a$ and $G_i$
is nothing but a generalized FPLD model with $n'=1$ and $(n-1)$, respectively.

One has now the right to update these generalized FPLD models via any valid Monte Carlo algorithm.
We choose {\it Algorithm 2} to update the model with $n'=1$ on $G_a$ and the identity operation (`do nothing')
on $G_i$. Due to the fact $n'=1$, the loop-number change $\Delta \ell$ does not matter anymore.
Therefore, one can formulate another version of the worm algorithm as ({\it Algorithm 3})
\begin{enumerate}
  \item Do the {\it Coloring assignment} if $u=v$.
  \item Do $M$ times of the coarse Monte Carlo steps (from and back to the M sector)
           by performing {\it Algorithm 2} on the induced subgraph $G_a$ with $n'=1$.
\end{enumerate}
The parameter $M \geq 1$ can be set such that step 1 and 2 take comparable CPU time.

For the actual implementation of {\it Algorithm 2} and {\it 3}, positive statistical weights
have to be assigned to vertex states in Fig.~\ref{fp_ld_W}. Before discussing on this,
we mention that there exist some freedom to choose which vertex state is allowed in the W sector.
As long as ergodicity is satisfied, the consideration is to optimize the efficiency.
In Fig.~\ref{fp_ld_W}, we do not allow the state with two bonds and a dimer,
because the only way to generate this state is to add a dimer to
state-$7$ and the only way to return to Fig.~\ref{fp_ld} is to delete the newly generated dimer.
Thus, such a state will not help updating the configuration while increasing computational burden.
In contrast, state-$9'$ and -$10'$ (-$11'$ and -$13'$) are important for moving around the dimers (bonds).
We set
\begin{eqnarray}
  W_{9'}  &=& W_{10'} = 1 \hspace{2mm}   W_{12'} = \min (1,W_7) \hspace{2mm} \mbox{ and} \nonumber \\
  W_{11'} &=& W_{13'} = W_7 \;\; = e^{-2K} \; .
\end{eqnarray}
State-$12'$ is useful for switching between dimer and bond, but should not occur more
frequently than state-6 or -7.
\section{Results}
\label{results}
\begin{figure}
\includegraphics[width=7cm,height=6.3cm]{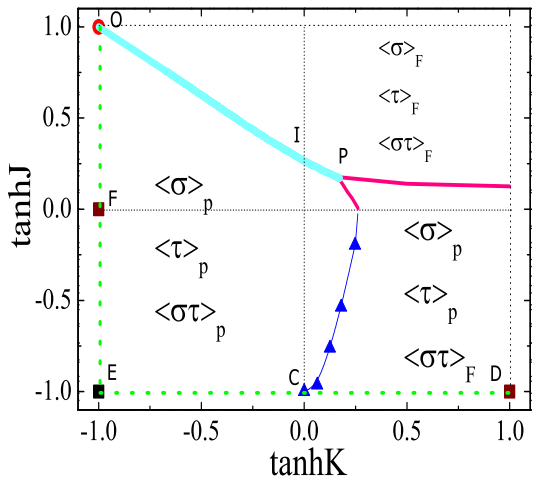}
\caption{\label{PD} (Color online) Phase diagram of the
AT model on the triangular lattice. Points--`P', `I', `O'--correspond to
the 4-state Potts, the Ising, and the O(2) loop model, respectively.
Points--`F', `C', `D'--denote the zero-temperature Ising antiferromagnet in
variable $\sigma \tau$, $\sigma$ or $\tau$ (decoupled), $\sigma$ or $\tau$ (correlated),
respectively. Point `E' is the zero-temperature 4-state Potts antiferromagnet.}
\end{figure}

The complete phase diagram of AT model on the triangular lattice is shown in Fig.~\ref{PD}.
In following, we shall present numerical results and discuss the phase boundary in the
antiferromagnetic two-spin coupling region ($J<0$).

\subsection{Finite $J$}
We employ {\em Algorithm 1} with Eq.~(\ref{vertex_weight_W1}) to simulate the AT
model in the region of finite $J<0$ on triangular lattices with periodic boundary conditions,
using system sizes in the range $6 \leq L \leq 192$.

For a given loop configuration, we generate the associated spin configuration
on the triangular lattice according to the low-temperature expansion rule.
Note that, due to the periodic boundary condition, a loop configuration
may correspond to no spin configuration. This occurs when there exists
an odd number of red or blue loops winding around the boundary. In this case,
we take no measurement, and the simulation continues until the next try.
Let $X$ be the indicator function which is $1$ if the loop configuration is measuring and
corresponds to a spin configuration, and $0$ otherwise; let $\mathcal{O}$ be the operator computed in
one of the $4$ compatible spin configurations; therefore, what we are computing is $[ \mathcal{O} X ]/[ X ]$, with $[ \; ]$ the statistical average
over loop configurations. The non-valid loop configuration is not weighted,  thus it does not influence any of the numerical data related to the
spin variables. Further, since the special cases that the loop configuration does not correspond to any spin configuration
result from boundary effects, such cases do not dominate in large systems.

Two types of magnetization are measured as
\begin{equation}\label{M}
M_{\sigma}=\frac{1}{V}\sum_{i}  \sigma_{i}
\hspace{5mm} \mbox{and} \hspace{5mm}
M_{\sigma \tau}=\frac{1}{V} \sum_{i} \sigma_{i} \tau_{i} \; ,
\end{equation}
where the summation is over the whole lattice.
Accordingly, the susceptibilities are defined as
\begin{equation}
  \chi_{\sigma}= V \langle M_{\sigma}^{2} \rangle
  \hspace{5mm} \mbox{and} \hspace{5mm}
  \chi_{\sigma \tau}=V \langle M_{\sigma \tau}^{2} \rangle \;,
\end{equation}
with $\langle \; \rangle$ for statistical average.
Dimensionless ratios are found to be very powerful in locating the critical points
of many systems under continuous phase transitions. On the basis of
the fluctuation of the magnetization,
we define two distinct dimensionless ratios as ~\cite{binder}
\begin{equation}
  Q_{\sigma} = \frac{\langle M_{\sigma}^{2} \rangle^2}{\langle M_{\sigma}^{4} \rangle}
\hspace{3mm} \mbox{and} \hspace{3mm}
Q_{\sigma \tau}= \frac{\langle M_{\sigma \tau}^2 \rangle^2}{\langle M_{\sigma \tau}^4 \rangle} \; .
\end{equation}
We also measure energy-like quantities as
 \begin{eqnarray}
 E_{\sigma}    &=& -J \sum_{\langle i,j\rangle }  \sigma_i \sigma_j \label{Ham22} \\
 E_{\sigma\tau} &=& -K\sum_{\langle i,j\rangle }  \sigma_i \tau_i \sigma_j \tau_j  \label{Ham23} \\
 E &=& E_{\sigma}+ E_{\tau} + E_{\sigma\tau} \label{Ham24} \; ,
\end{eqnarray}
as well as the associated specific-heat-like quantities $C_\sigma=(\langle E_\sigma^{2}
\rangle -\langle E_\sigma \rangle ^{2})/V$, $C_{\sigma \tau}$, and $C$.
\begin{figure}
\includegraphics[width=7cm,height=5cm]{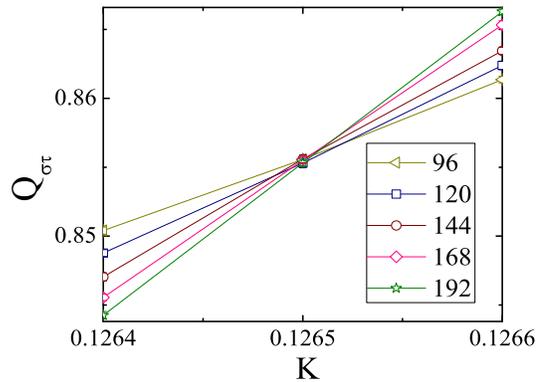}
\caption{\label{bdj10} (Color online) Quantity
$Q_{\sigma\tau}$ versus $K$ at $J=-1.0$. Lines connecting the data points
are for illustration purpose.}
\end{figure}

The AT model for $J=0$ reduces to the standard Ising model in the Ising-spin variable $\sigma \tau$,
and undergoes a Ising-like transition at $K_c$.
For $K < K_c$, the configurations in the Ising variables $\sigma$, $\tau$,
and $\sigma \tau$ are all in the disordered (paramagnetic) state;
for $K >K_c$, $\sigma \tau$ is in the ferromagnetic state while $\sigma$
and $\tau$ are still in the paramagnetic state. We expect that this scenario
continues into the region $J <0$.

We choose $J=-0.2, -0.6, -1.0$, and $-2.0$, and perform some preliminary and
coarse simulations to approximately locate the intersection of  $Q_{\sigma \tau}$ for various linear
system sizes $L$.
Then, fine and extensive simulations are carried out near the estimated critical point.
Figure~\ref{bdj10} displays $Q_{\sigma \tau}$ versus $K$ for different $L$
at $J=-1.0$, indicating a critical point near $K \approx 0.1265$.

The finite-size scaling behavior of $Q_{\sigma \tau}(K,L)$ near the critical point $K_c$ is described by
\begin{equation}
Q(K,L) = Q(tL^{y_t}, b L^{y_i}) \; ,
\label{scaling_Q}
\end{equation}
where $t$ and $i$ represent the leading and the subleading thermal scaling fields, with $t \propto
(K-K_c) + \cdots$. The associated renormalization exponents are denoted as $y_t$ and $y_i$.
A Taylor expansion of Eq.~(\ref{scaling_Q}) yields~\cite{Deng}
\begin{eqnarray}
Q(K,L)&=&  Q_c+a_1 \Delta K L^{y_t} +a_2(\Delta K)^2 L^{2y_t}+ b L^{y_i} \nonumber \\
      &+&  c \Delta K L^{y_{t}+y_{i}}+... \; ,
\label{fit}
\end{eqnarray}
with $\Delta K \equiv K-K_c$. Parameters $a_{1}$, $a_{2}$, $b$, and $c$
are unknown constants.

According to the least-squares criterion, we fit the $Q_{\sigma \tau}$ data to Eq.~(\ref{fit}).
Assuming the transition is Ising-like, we expect that the leading two finite-size correction
exponents are $y_1 = 2-2y_h =-7/4$ and $y_2=y_i=-2$ for $Q_{\sigma \tau}$,
where $y_h=15/8$ is the magnetic renormalization exponent.
With $y_1$ and $y_2$ fixed and $L \geq L_{\rm min} = 48$,
we obtain $K_c=0.12653 (2)$, $y_t=1.01 (2)$, and $Q_c=0.8587 (1)$ for $J=-1.0$. The chi square per
degree of freedom ($\bar{\chi}^2$/dof) is 1.14.
The estimate of $y_t$ is consistent with the
exact result  $y_t=1$, and the universal ratio $Q_c=0.8587$ also agrees
well with the earlier estimate $Q_c=0.858\, 725 \, 28 (3)$ for the Ising model
on the triangular lattice~\cite{blote1}.

The data of susceptibility $\chi_{\sigma\tau}$ is analyzed by
\begin{eqnarray}
\chi(K,L)&= & L^{-2y_h+d}( a_0+a_1 \Delta K L^{y_t} +a_2(\Delta K)^2 L^{2y_t}  \nonumber \\
      &+&  b L^{y_{i}} + c \Delta K L^{y_t+y_i}+...)  \; ,
\label{fit2}
\end{eqnarray}
and we determine the magnetic exponent as $y_h=1.876(2)$, in good agreement with
the exact value $y_h=15/8$. The specific-heat-like quantity $C$ is also found to diverge
approximately in the logarithmic scale as $L$ increases.
No phase transition is observed for Ising variable $\sigma$ or $\tau$.

Similar results are found for other values of $J$, and the estimated critical
points are listed in Table~\ref{tb5}.

\begin{table}
\begin{center}
\begin{tabular}{|c|cccc|}
  \hline
  $J$ & -0.2               & -0.6             & -1.0              & -2.0    \\
  \hline
$K_c$ & $0.25303 \,(2)$   & $0.18164 \,(2)$ & $0.12653 \,(2)$  & $0.06306 \,(3)$  \\
  \hline
$L_{\rm min}$ & 48   & 48 & 48  & 48  \\
  \hline
$\bar{\chi}^2$/dof & 1.05  & 0.86 & 1.14  & 1.21  \\
  \hline
\end{tabular}
\caption{Details in the data fits according to Eq.~\ref{fit}.}
\label{tb5}
\end{center}
\end{table}

On this basis, we conclude that the phase transition of the AT model in region $(K>0,J<0)$ with finite $J$
is in the Ising universality. Finally, we mention that
the worm-type algorithm hereby does not suffer much from critical slowing-down.

\begin{figure}
\includegraphics[width=7cm,height=5cm]{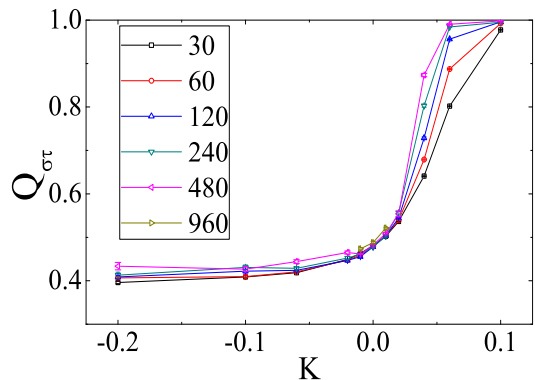}
\caption{\label{sbbd} (Color online) Dimensionless ratio $Q_{\sigma\tau}$ versus $K$ for the n=2 FPLD model.}
\end{figure}

\subsection{$J \rightarrow -\infty$}

Table~\ref{tb5} suggests that the critical coupling $K_c$ becomes smaller as $J$ becomes more negative, and that
the ending point of the critical line for $J \rightarrow - \infty$ is very close to $K=0$, since
$K_c (J=-2) =0.06306 (3)$ is already near $0$.  To locate the ending point more accurately,
we directly simulate the $J \rightarrow - \infty$ limit, which makes use of the exact mapping
to the $n=2$ FPLD model and employs {\it Algorithm 3}. System sizes take $6$ values in range $30 \leq L \leq 960$.

Note that the loops in the FPLD model serve as domain walls for the Ising variable $\sigma \tau$
in the AT model. According to the low-temperature expansion rule, on the triangular lattice
we sample magnetization density $M_{\sigma \tau}$, susceptibility  $\chi_{\sigma \tau}$,
dimensionless ratio $Q_{\sigma \tau}$, energy $E_{\sigma \tau}$, and specific heat $C_{\sigma \tau}$,
whose definitions can be found in Eqs.~(\ref{M})--(\ref{Ham24}).
Further, to explore the loop-length distribution, on the honeycomb lattice
we measure the length of the longest loop as $S_1$.

The finite-size data of the dimensionless ratio $Q_{\sigma \tau}$
are plotted in Fig.~\ref{sbbd}; an eye-view fitting yields a critical
point as $K_c=0.00(2)$. For $K >K_c$, the $Q_{\sigma \tau}$ value rapidly approaches to $1$
as size $L$ increases.
This reflects that the Ising variable $\sigma \tau$ exhibits a long-range ferromagnetic order
on the triangular lattice; correspondingly, on the honeycomb lattice loops are small--i.e.,
in a disordered state. For $K<K_c$,  $Q_{\sigma \tau}$ converges to a constant $Q_c$
which deviates from the trivial Gaussian value $1/3$. This implies that, despite the absence of
a long-range order, the spin-spin correlation function decays algebraically over the distance.

\begin{figure}
\includegraphics[width=7cm,height=5cm]{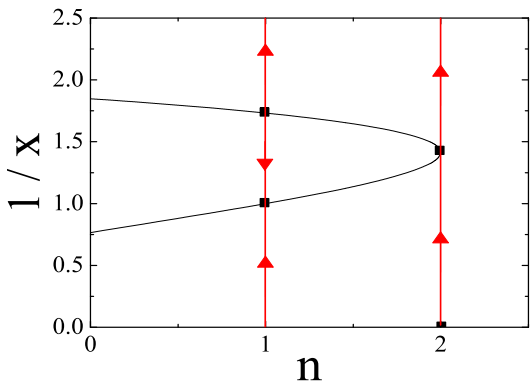}
\caption{\label{lpcp} (Color online) Phase diagram of the O(n) loop model~\cite{Nienhuis}.
Red lines denote the directions of the renormalization flows.}
\end{figure}

In Fig.~\ref{sbbd}, one can observe that $Q_{\sigma \tau}$ at $K<0$ rapidly converges to a K-dependent value,
as expected in the low-temperature BKT phase. This reminds us the analogy between the FPLD and
Nienhuis's O$(n)$ honeycomb loop model with $n=2$.  The phase diagram of the latter is shown in
Fig.~\ref{lpcp}, where $x$ is the statistical weight for an occupied bond. For a given $0 \leq n \leq 2$,
the O$(n)$ loop model exhibits three distinct phases: a dilute and disordered phase (small $x$),
a densely-packed phase (large finite $x$), and a fully-packed phase (infinite $x$). Furthermore,
the model is exactly solvable on the curves~\cite{Nienhuis}
\begin{equation}
  \frac{1}{x_{\pm}} = \sqrt{2 \pm \sqrt{2-n} } \; .
\end{equation}
The system is equivalent to the {\em tricritical} $q=n^2$ Potts model along the critical line $x_+$,
belongs to the {\em critical}  $q=n^2$ Potts universality class in the densely-packed phase,
and is in another critical universality in the fully-packed phase.
For $n=2$, the two solvable lines $x_{\pm}$ merge at a single point;
the renormalization field is {\it marginally} relevant (irrelevant) for $x<x_{\pm}$ ($x > x_{\pm}$).
In other words, the phase transition at $x_{\pm}$ is Berezinsky-Kosterlitz-Thouless(BKT)-like.
At the special point $x_{\pm} (n=2)$, the amplitude of the renormalization field is zero,
and thus logarithmic corrections, present at most of BKT-like critical points, disappear.
This explains the absence of logarithmic corrections in the critical Baxter-Wu model,
which can be exactly mapped onto the O(2) loop model at $x_{\pm}$.
For the critical O(2) loop model, it has been identified that $S_1 \propto L^{y_H} = L^{3/2}$ and
$\chi_{\sigma \tau} \propto L^{2y_{t0}-2}=L$, where $y_H = 3/2$ is the hull exponent
and $y_{t0}=3/2$ is the leading thermal renormalization exponent in the language of the Potts model~\cite{dengy07}.

\begin{figure}
\includegraphics[width=7cm,height=5cm]{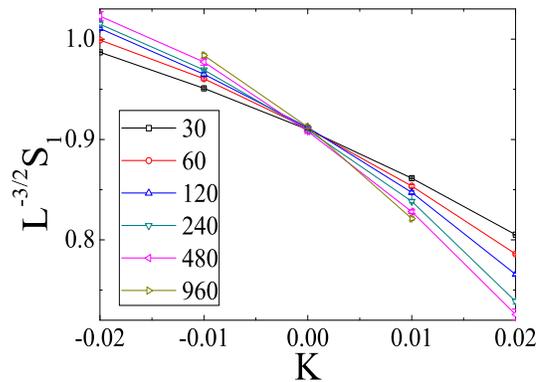}
\caption{\label{S1} (Color online) $L^{-3/2}S_1$ versus $K$ for the n=2 FPLD model.}
\end{figure}

\begin{figure}
\includegraphics[width=7cm,height=5cm]{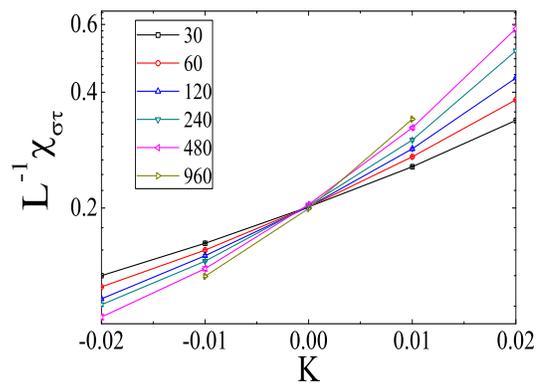}
\caption{\label{sus1} (Color online)
$L^{-1}\chi_{\sigma\tau}$ versus $K$ for the n=2 FPLD model.}
\end{figure}

Since the state space of the FPLD model is a subspace of the O(2) loop model, it is reasonable to
conjecture that the two models are in the same universality class.
Namely, we expect that the FPLD model undergoes
a BKT-like transition at $K_c$, where the logarithmic corrections are absent; for $K<K_c$
the system is in the same universality class as at $K_c$ but with logarithmic corrections;
for $K \rightarrow - \infty$ it is in another universality class. Making use of the known
exponent $y_H = 3/2$ for $S_1$ and $2y_{t0}-2=1$ for $\chi_{\sigma \tau}$, we plot $ L^{-3/2} S_1$
and $L^{-1} \chi_{\sigma \tau}$ versus $K$ in Figs.~\ref{S1} and \ref{sus1}, respectively.
They both display a nice intersection at $K=0.000$.
From Fig.~\ref{sus1} one can observe that the exponent $y_H$ varies along the BKT critical line, which reconciles the
difference of $y_H$ between the present model and the two-dimensional XY models.

\begin{figure}
\includegraphics[width=7cm,height=5cm]{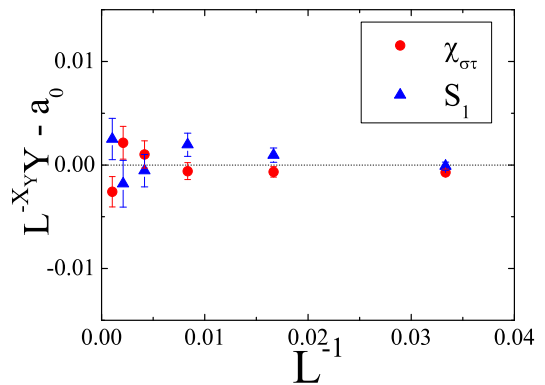}
\caption{\label{k0} (Color online) Quantities $L^{-3/2} S_1-a_{0,s_1}$
and $L^{-1} \chi_{\sigma \tau}-a_{0,\chi}$ at $K=0$ versus $L^{-1}$ for the n=2 FPLD model.
Constants $a_{0,s_1}$ and $a_{0,\chi}$ are obtained from the fits.}
\end{figure}

To further explore the potential logarithmic corrections, we assume $K_c=0$ and
plot $L^{-3/2} S_1$ and $L^{-1} \chi_{\sigma \tau}$ at $K=0$ versus $L^{-1}$.
As shown in Fig.~\ref{k0}, the rapid convergence implies the absence of logarithmic
corrections; corrections with term $L^{-1}$ are also very weak if they exist.

According to the least-squares criterion, the $S_1$ and $\chi_{\sigma \tau}$ data are fitted by
\begin{eqnarray}
  && Y(K,L) = c_0 + c_1 \Delta K +\cdots + L^{X_Y}(a_0+a_1 \Delta K \ln L \nonumber \\
  && + a_2(\Delta K)^2 \ln^2L + b_1 L^{y_1}+b_2 L^{y_2}+...) \; .
  \label{fit22}
\end{eqnarray}
Here $a_i$ are coefficients of the finite-size scaling variable $\Delta K \ln L$ with
$\Delta K = K- K_c$, $b_i$ are amplitudes of finite-size corrections, and the
terms with $c_i$ accounts for analytical background. There are also cross-terms involving
products of terms arising from these three sources. The exponent $X_Y$ is a general label
for quantity $Y$. Equation~(\ref{fit22}) has assumed the absence of logarithmic corrections.
It occurs that both the $S_1$ and the $\chi_{\sigma \tau}$ data with $L \geq L_{\rm min}= 60$
can be well described by Eq.~(\ref{fit22}) with fixed correction exponents $y_1 =-1$ and $y_2=-2$.
The results for $S_1$ are $X_Y = 1.498 (3)$, $K_c = 0.001 (2)$, and $\bar{\chi}^2$/dof=1.22; for
$\chi_{\sigma \tau}$ are $X_Y = 1.001 (2)$, $K_c =-0.001 (1)$, and $\bar{\chi}^2$/dof=0.87. These agree well
with the known exponents $3/2$ for $S_1$ and $1$ for $\chi_{\sigma \tau}$, as well as
with the expectation $K_c =0 $. If the exponents $X_Y$ are further fixed at the known values,
we obtain $K_c = 0.0002 (3)$, $\bar{\chi}^2$/dof=0.94 from $S_1$ and  $-0.0003 (4)$,  $\bar{\chi}^2$/dof=1.09
from $\chi_{\sigma \tau}$. On this basis, we estimate the critical point as $K_c = -0.0001 (6)$,
which covers the uncertainties of $K_c$ from $S_1$ and $\chi_{\sigma \tau}$.

We mention that, when an external field of strength $h/T$ is applied to the triangular Ising
antiferromagnet, the critical state of the system is
not immediately destroyed. Instead, the system has a BKT-like transition
at $h_{c}=0.266(10)$ ~\cite{blote2}.
However, our Monte Carlo results suggest that a critical
point $K_c \neq 0$ is rather unlikely for the $n=2$ FPLD model.

In the limit $ K \rightarrow \infty$, the Ising variable $\sigma \tau$ is in the ferromagnetic state.
However, in terms of the $\sigma$ or the $\tau$ variable, it can be easily derived that
the system is also an Ising model with coupling $2J$. Namely, along the $\tanh K =1$ line, the
AT model has an Ising-like transition at $\tanh 2J = \sqrt{2} -1$. Further, the corner point
${\rm D} : \equiv (\tanh K=1, \tanh J = -1)$ corresponds to the triangular antiferromagnet at zero temperature,
which is critical.
Together with the earlier discussions in Sec. I, this means that, in Fig.~\ref{PD},
the limiting points--D, C, O, F, E--are all critical.
From our simulations in range $-0.2 \leq K \leq 0.1$ along the $\tanh J =-1$ line (EC+CD),
we observe that, in the whole range, there exist algebraically decaying two-point
correlation function for the $\sigma$ or the $\tau$ variable. On this basis, we conjecture that
the whole $\tanh J = -1$ line (EC+CD) is critical for the $\sigma$ or the $\tau$ variable.
Simulation along the $\tanh K = -1$ line using the present worm algorithms suffers significantly
from critical slowing-down. Nevertheless, we suspect that the whole  $\tanh K = -1$ line
is critical for the $\sigma \tau$ variable.

\section{Dynamic Critical Behavior}
\label{worm dynamic data}

In this section, we briefly report the efficiency of  {\it Algorithm 2} for the $n=2$ FPLD model,
using the standard procedure described in Ref.~\cite{SokalLectures}.

For each observable (say $\mathcal{O}$), we calculate its autocorrelation function
\begin{equation*}
\rho_{\mathcal{O}}(t)= \langle \mathcal{O}(t)\mathcal{O}(0)\rangle - \langle \mathcal{O}\rangle^2,
\end{equation*}
where $\langle$ $\space$ $\rangle$ denotes expectation with respect to the stationary distribution.
We then obtain the corresponding integrated autocorrelation time as
\begin{equation}
   \tau_{{\rm int},\mathcal{O}}= \frac{1}{2}\, \sum_{t = -\infty}^{\infty} \rho_{\mathcal{O}}(t) \;.
\end{equation}

The dynamic critical exponent $z_{{\rm int},\mathcal{O}}$ is defined by
\begin{equation}
\label{tau definitions}
\tau_{{\rm int},\mathcal{O}}
\sim \xi^{z_{{\rm int},\mathcal{O}}}.
\end{equation}
where $\xi$ is the spatial correlation length.
On a finite lattice at criticality, $\xi$ is cut off by system size $L$.
Therefore, one has
\begin{equation}
\tau_{{\rm int},\mathcal{O}} = a + b L^{z_{{\rm int},\mathcal{O}}},
\label{fit_tau_int}
\end{equation}
with $a$ and $b$ unknown parameters.

\begin{figure}
\includegraphics[width=7cm,height=5cm]{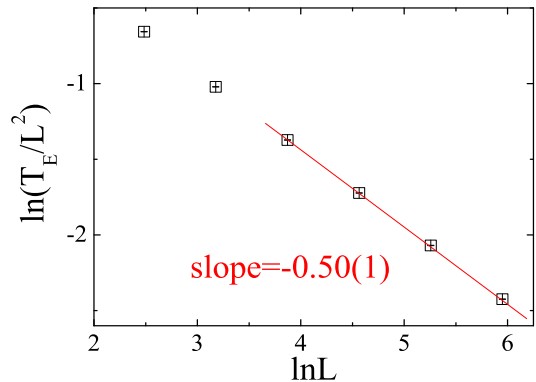}
\caption{\label{Te} (Color online) Ln($T_E/L^2$) versus lnL at $K=0$.  }
\end{figure}

We simulate at the critical point $K_c=0$.
Note that, during the worm simulations we measure the observables
only when the chain visits the Eulerian subspace, roughly every $T_E\sim L^{d-2X_{e}}$ hits. However, it is natural to define $z_{\text{int},\mathcal{O}}$ as in Ref.~\cite{improved} to measure time in units of
{\em sweeps} of the lattice, i.e. $L^d$ hits. Since one sweep takes of order $L^{2X_{e}}$ visits to the Eulerian subspace,
we have $\tau\sim L^{z+2X_{e}}$. As shown in Fig.\ref{Te}, the exponent $2X_{e}$ is estimated to be $0.50(1)$.

\begin{figure}
\includegraphics[width=7cm,height=5cm]{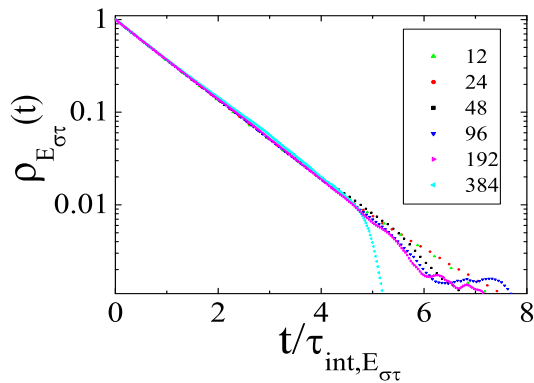}
\caption{\label{rou_D} (Color online) $\rho_{E_{\sigma \tau}}(t/\tau_{int,E_{\sigma \tau}})$ versus $t/\tau_{ {\rm int},
E_{\sigma \tau}}$ at $K=0$.}
\end{figure}

\begin{figure}
\includegraphics[width=7cm,height=5cm]{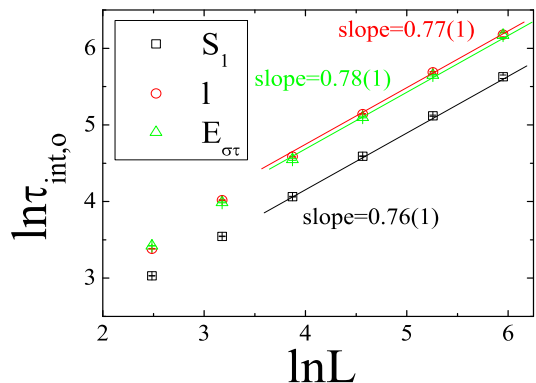}
\caption{\label{tau} (Color online) Ln$(\tau_{{\rm int},\mathcal{O}})$ versus lnL for different observables at $K=0$. }
\end{figure}

Among the measured quantities including the longest-loop length $S_1$, the
loop number $\ell$, and the energy-like quantity $E_{\sigma \tau}$ etc,
$E_{\sigma \tau}$ is found to have the largest value of $\tau_{\rm int}$.
Figure~\ref{rou_D} displays $\rho_{E_{\sigma \tau}}(t/\tau_{ {\rm int},E_{\sigma \tau}})$
as a function of $t/\tau_{int,E_{\sigma \tau}}$, where an approximately exponential decay is observed.
The $\tau_{int,E_{\sigma \tau}}$ data are analyzed, and
we obtain  $z_{ {\rm int},E_{\sigma \tau}}=0.28(1)$, which is shown in Fig.~\ref{tau}. Similar fits are done for other quantities,
and we have  $z_{ {\rm int},S_1}= 0.26(1)$ and $z_{ {\rm int},\ell}= 0.27(1)$. In these fits, $\bar{\chi}^2$/dof ranges from $0.74$ to $1.31$.
Therefore, our numerical results suggest that the present worm algorithm is even more efficient than the one in Ref.~\cite{improved}.

Simulations are also carried out for $K=-0.05$, and the dynamic critical behavior cannot be
distinguished from that for $K=0$.

\section{Discussion}\label{discussion}

In summary, we have formulated two versions of the worm-type algorithms for the AT model
on the triangular lattice. The algorithms are based on the low-temperature expansion graph of the AT model,
and use the language of vertex states. Such a formulation not only provides
us a different angle to understand the worm method, but also offers
an easy way to overcome the absorbing difficulty.
The efficiency of our algorithm is studied and can also be reflected by the fact that
we can simulate up to size $L=960$.  Apparently, {\it Algorithm 1} can be applied to
the ferromagnetic region $J>0$ of the triangular AT model and to the AT model on other planar lattices.
Further, we mention that the worm-type algorithms can be developed on the basis of
the high-temperature expansion graph of the AT model. This yields a graphical model also
by Eq.~(\ref{partition_lt1}), but defined on the original lattice for the AT model. The
statistical weights of the occupied bonds are
\begin{eqnarray}
  X_r&=&X_b=(e^{2K} \sinh 2J)/(e^{2K} \cosh 2J +1) \nonumber \\
  X_{r+b}&=&(e^{2K} \cosh 2J-1)/(e^{2K} \cosh 2J +1) \; .
\end{eqnarray}
It is reasonable to expect good efficiency for the AT model on non-planar lattices--e.g., in
higher spatial dimensions--with non-negative weights $X_r=X_b$ and $X_{r+b}$.

The high efficiency of the worm algorithms allows us to explore the triangular-lattice AT model
in the antiferromagnetic region, and accordingly we conjecture a complete phase diagram in
the $(J,K)$ plane. Of the particular interest is the $J \rightarrow - \infty$ limit, where
the AT model is mapped onto the FPLD model with $n=2$. As suggested by the Monte Carlo simulation,
the AT model undergoes a BKT-like transition along the $\tanh J = -1$ line, in the same
universality class as the classical $XY$ model. We also mention that it remains to be
explored whether or not, for other values of $n$, the FPLD and Nienhuis's O$(n)$ model are
in the same universality class.

\begin{figure} \includegraphics[width=7cm,height=6.3cm]{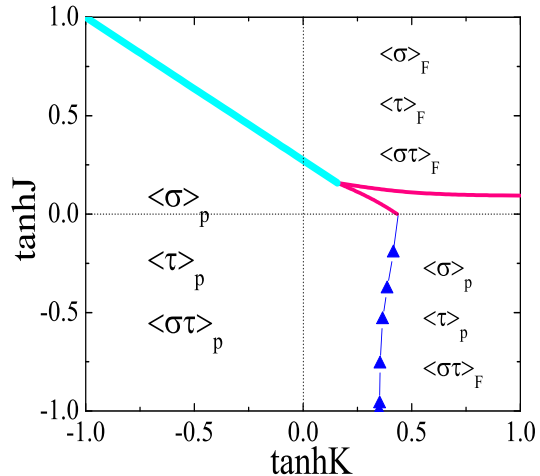}
\caption{\label{PDKG} (Color online) Phase diagram of the AT model
on the ka\'gome lattice.}
\end{figure}

Finally, we perform simulations for the AT model on the ka\'gome lattice in the
region $(J<0, K \geq 0)$, and determine a line of Ising-like critical points.
The results are shown in Table~\ref{tb3}. Unlike on the triangular lattice,
the critical line ends at $K_c = 0.3655 >0$, still in the Ising universality.
Taking into account that the frustration on the ka\'gome lattice is only partial,
this is not surprising.
Accordingly, the phase diagram is shown in Fig.~\ref{PDKG}.

\begin{table}
\begin{center}
\begin{tabular}{|c|cccccc|}
  \hline
  $J$ & -0.2              & -0.4   & -0.6             & -1.0       & -2.0  & $-\infty$  \\
  \hline
$K_c$ & $0.442(2)$  &$0.408(2)$ & $0.386(2)$ & $0.370(2) $  & $0.366(2)$ & 0.3655(3) \\
  \hline
$L_{min}$ & 36   & 36 & 36  & 36 & 36 & 36  \\
  \hline
$\bar{\chi}^2$/dof & 1.21  & 0.91 & 1.07  & 1.23 & 0.85 & 1.15 \\
  \hline
\end{tabular}
\caption{Details in the data fits according to Eq.~\ref{fit} on the ka\'gome lattice.}
\label{tb3}
\end{center}
\end{table}

\section{Acknowledgements}
The work of Q.H.C was supported by National Basic Research Program of China (Grant Nos. 2011CBA00103 and 2009CB929104). 
The work of Y.D was supported by NSFC (Grant No. 10975127),
Anhui Provincial Natural Science Foundation (Grant No. 090416224) and CAS.

\end{document}